# Adversarial Fragility and Language Vulnerability in Clinical AI:

## A Systematic Audit of Diagnostic Collapse Under Imperceptible Perturbations and Cross-Lingual Drift in Low-Resource Healthcare Settings

**Anthonio Oladimeji Gabriel**[1,*] **and Ahmad Rufai Yusuf**[2]

*1 Centre for Clinical Intelligence & Safety, Ìyàwó, Nigeria*

*2 Tomorrow University of Applied Sciences, Germany*

***Corresponding author: anthoniooladimeji11@gmail.com***

## Abstract

Current clinical artificial intelligence (AI) systems are evaluated almost exclusively on clean, standardised, English-language inputs - conditions that do not reflect the realities of healthcare delivery in low-resource settings. This study presents the first systematic dual audit of two orthogonal safety vulnerabilities in clinical AI: adversarial image fragility and cross-lingual diagnostic drift. Using DenseNet121 - the architecture underlying CheXNet - fine-tuned on the COVID-QU-Ex chest X-ray dataset (85,318 images; COVID-19, Non-COVID Pneumonia, Normal), we demonstrate that diagnostic accuracy collapses from 89.3% to 62.0% under a Fast Gradient Method (FGM) perturbation of ε=0.021 - a magnitude imperceptible to the human eye. Standard defensive strategies, including Gaussian smoothing and ensemble voting, failed to restore clinical safety. In a parallel language fragility experiment, we tested Llama3.1:8b and NatLAS (N-ATLAS) on 20 COVID-19 clinical cases presented in Standard English, Nigerian Pidgin (Naija), and Yoruba-inflected English. Both models exhibited significant accuracy degradation: Llama3.1:8b dropped from 80.0% to 65.0% on Pidgin; NatLAS - an African-context model - collapsed from 85.0% to 55.0%, with diagnosis consistency falling to 50%. These findings establish a quantitative failure envelope for clinical AI under conditions representative of Primary Health Centre (PHC) deployment in Nigeria, and motivate urgent calls for adversarially hardened, linguistically inclusive clinical AI architectures.

## 1. Introduction

Sub-Saharan Africa carries approximately 24% of the global disease burden while commanding less than 1% of global health expenditure [1]. In this context, artificial intelligence has emerged as a promising force multiplier for diagnostic capacity. Smartphone-based AI algorithms have demonstrated 98.5% accuracy in identifying Plasmodium falciparum in thick blood smears in Kenya, reducing inappropriate antibiotic prescribing by 31% [1]. In West Africa, AI-assisted chest radiograph interpretation has reduced diagnostic turnaround times by approximately 40% in radiologist-scarce facilities [1]. Nigeria, with nearly 240 million people and only one pathologist per 500,000 population, has rapidly adopted computer-aided detection (CAD) systems [1].

However, the transition from retrospective laboratory validation to real-time clinical application in Nigerian Primary Healthcare Centres (PHCs) is hindered by a fundamental 'fragility envelope' in current deep learning architectures [2]. Standard evaluations are conducted on clean, standardised, and English-dominant datasets - conditions that do not reflect the linguistic diversity, imaging variability, or latent adversarial threats of real-world PHC deployment.

This study addresses two critical and previously unaudited vulnerabilities: (1) the susceptibility of clinical image AI to imperceptible adversarial perturbations, and (2) the diagnostic drift of medical large language models (LLMs) when patients present in Nigerian Pidgin (Naija) or Yoruba-inflected English rather than Standard English. Both vulnerabilities are directly relevant to the Ìyàwó Clinical Decision Support System, which serves Community Health Extension Workers (CHEWs) across 264 PHCs in Oyo State, Nigeria.

Our contributions are: (i) **the first quantitative adversarial robustness audit of DenseNet121 on COVID-19 chest X-rays in a Nigerian clinical framing**; (ii) **a systematic cross-lingual diagnostic drift evaluation across three Nigerian language registers;** (iii) **a head-to-head comparison of a general-purpose LLM (Llama3.1:8b) and an African-context model (NatLAS) on clinical language fragility;** and (iv) **empirical evidence that standard defensive strategies fail to restore clinical safety under these conditions**.

## 2. Background and Related Work

### 2.1 Adversarial Vulnerability in Medical Imaging

DenseNet121 is favoured for chest X-ray classification due to its dense connectivity mechanism, which mitigates the vanishing gradient problem by allowing each layer to receive feature maps from all preceding layers [3]. This architecture underpins CheXNet, which achieved radiologist-level performance across 14 pulmonary diseases [4]. However, the dense connectivity that enhances feature extraction also creates a complex gradient landscape susceptible to adversarial manipulation [3].

Gradient-based evasion attacks such as the Fast Gradient Sign Method (FGSM) exploit the model's own loss gradients to construct adversarial examples: $x_{adv} = x + \varepsilon \cdot \mathrm{sign}(\nabla_x J(\theta, x, y))$. Medical images exhibit higher sensitivity to these perturbations than natural images, hypothesised to occur because clinical diagnostics rely on subtle biological textures and minute intensity variations – features easily disrupted by small $\varepsilon$ values [8]. Comparative studies across architectures have found that standard-trained models are universally vulnerable, but modern transformer-inspired designs such as ConvNeXt demonstrate greater robustness through larger kernels and global dependency modelling [7]. Capsule Networks (CapsNets) have also emerged as robust alternatives, maintaining more consistent Grad-CAM attention maps after perturbation by modelling hierarchical spatial relationships rather than pixel patterns [8].

### 2.2 Language Fragility in Clinical LLMs

Over 90% of current LLM training data originates from Standard American English (SAE), with less than 7% from other languages and negligible representation from West African dialects [16]. Nigerian Pidgin (Naija) - spoken by approximately 120 million people - is systematically underrepresented [14]. Research using the WARRI benchmark dataset has demonstrated that LLMs adapted to West African Pidgin English (WAPE, the BBC variety) achieve ChrF++ scores of 76.3-83.4, while models processing community-level Naija score approximately 54 - a 36-point drop from SAE [14]. This 'Standard American English bias' causes LLM clinical reasoning to drift from accuracy to linguistic guesswork when inputs depart from the training distribution [14].

N-ATLAS (NatLAS), unveiled in September 2025 as an open-source model supporting Yoruba, Hausa, Igbo, and Nigerian-accented English, represents a national commitment to linguistic inclusion [18]. Trained on the WAXAL dataset comprising 11,000 hours of speech from 21 Sub-Saharan African languages [19], it nonetheless may suffer from insufficient medical annotations in local dialects - causing its clinical reasoning to remain 'tethered' to its English-language knowledge base [20].

## 3. Methods

### 3.1 Dataset

We utilised the COVID-QU-Ex dataset [4], comprising 85,318 chest X-ray images across three classes: COVID-19, Non-COVID Pneumonia, and Normal. The dataset provides pre-defined Train/Validation/Test splits with lung masks and infection segmentation annotations. For all experiments, we sampled 50 images per class from the Test split (N=150 total), stratified by class, using a fixed random seed (42) for reproducibility. Images were resized to 224×224 pixels and normalised using ImageNet statistics (mean=[0.485, 0.456, 0.406], std=[0.229, 0.224, 0.225]).

### 3.2 Model Architecture and Fine-Tuning

We selected DenseNet121 as the primary architecture, pre-trained on ImageNet (6,956,931 parameters). The final classification layer was replaced with a linear layer mapping to N=3 classes. Fine-tuning was performed on 200 images per class from the Training split (N=600 total) for 10 epochs using the Adam optimiser ($lr=1\times10^{-4}$) with StepLR scheduling (step_size=3, $\gamma=0.5$) and cross-entropy loss. All training utilised Apple Silicon MPS acceleration. Model weights were wrapped in the Adversarial Robustness Toolbox (ART) v1.20 PyTorchClassifier for attack generation.

### 3.3 Experiment 1: Adversarial Robustness Decay

We evaluated model accuracy across 15 epsilon levels linearly spaced between 0 and 0.3, using the Fast Gradient Method (FGM) attack. At each epsilon level, adversarial examples were generated for all 150 test images. Accuracy was computed as the proportion of correctly classified adversarial examples. Ninety-five percent confidence intervals were calculated using the Wilson score interval for proportions. Results were saved as experiment data and visualised as a robustness decay curve.

### 3.4 Experiment 2: Mitigation Stress Test

At the clinically critical threshold of $\varepsilon=0.021$ (the first attack point), we evaluated three defensive strategies: (1) Gaussian Smoothing - scipy gaussian_filter applied with $\sigma=1.0$ to adversarial images before classification; (2) Ensemble Voting - majority vote across five geometric augmentations (random horizontal flips and pixel shifts of $\pm5$); and (3) Mini Adversarial Training - five gradient steps of adversarial fine-tuning using Adam ($lr=1\times10^{-5}$) on batches of adversarial examples. Per-class accuracy and diagnosis consistency were computed for each condition.

### 3.5 Experiment 3: Language Fragility Audit

We constructed 20 COVID-19 clinical case vignettes covering COVID-19 (n=11), Non-COVID Pneumonia (n=5), and Normal (n=4) ground-truth diagnoses. Each case was authored in three registers: Standard English, Nigerian Pidgin (Naija), and Yoruba-inflected English - producing 60 unique query instances. Cases were designed to reflect realistic PHC triage presentations, including symptom descriptions, vital signs, and relevant exposure history. Each case was submitted to Llama3.1:8b and NatLAS (natlas:latest) via the Ollama local inference API (temperature=0.0 for reproducibility), with a structured prompt requesting a single diagnostic label from {COVID-19, Non-COVID Pneumonia, Normal}. Accuracy and cross-lingual consistency (agreement with Standard English prediction) were computed per model per language register.

## 4. Results

### 4.1 Fine-Tuning Performance

DenseNet121 achieved rapid convergence on the COVID-QU-Ex training subset, rising from 71.7% accuracy at Epoch 1 to 100.0% at Epoch 5, with training loss decreasing from 0.7201 to 0.0140 across 10 epochs (Figure 0). On the held-out test set, the fine-tuned model achieved an overall baseline accuracy of 89.3% (Figure 1). Per-class performance was: COVID-19 - 100.0% (precision=0.89, recall=1.00, F1=0.94); Non-COVID - 90.0% (precision=0.85, recall=0.90, F1=0.87); Normal - 78.0% (precision=0.95, recall=0.78, F1=0.86).

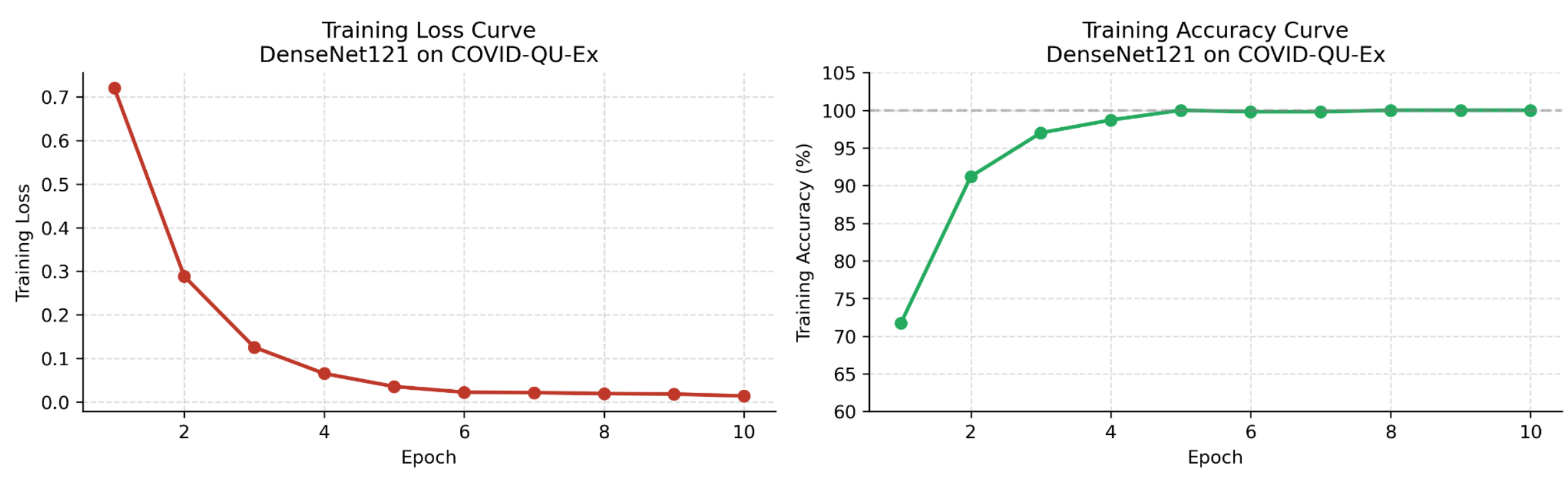

*Figure 0. Fine-tuning profile. Training loss (left) decreased from 0.7201 to 0.0140 across 10 epochs. Training accuracy (right) rose from 71.7% to 100.0%, confirming stable convergence.*

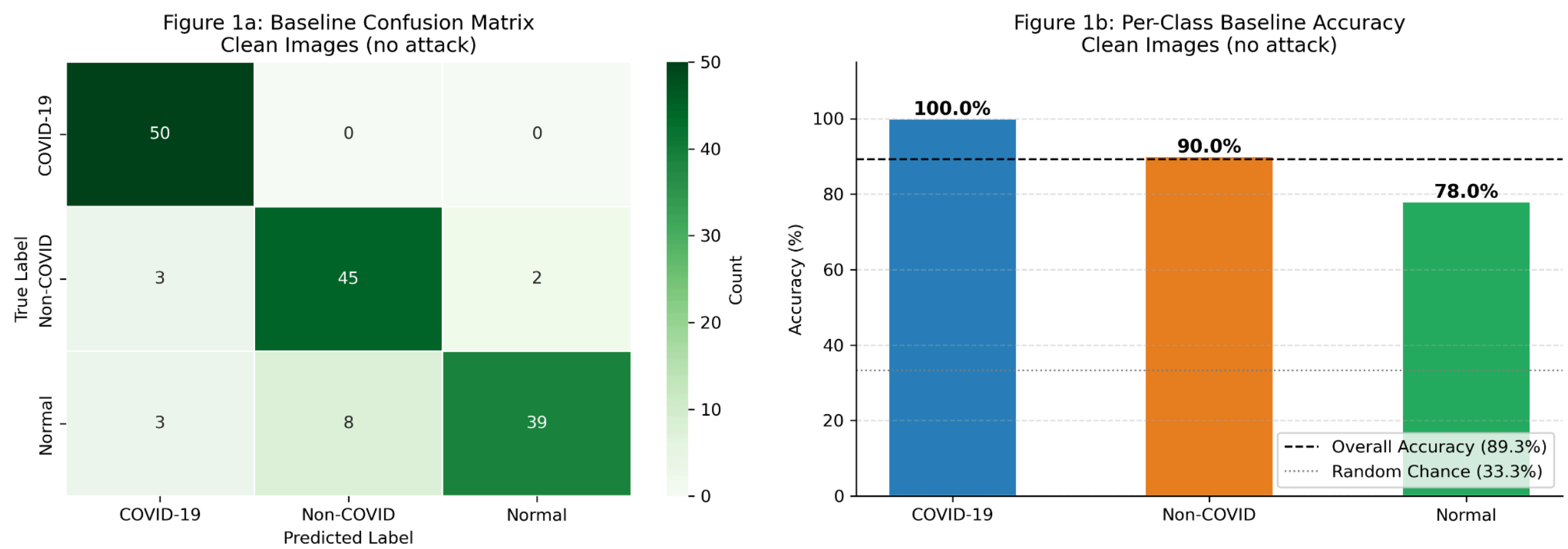

*Figure 1. Baseline performance on clean test images. (a) Confusion matrix showing 50/50 COVID-19 detection, with primary misclassifications occurring in the Normal class. (b) Per-class accuracy: COVID-19=100%, Non-COVID=90%, Normal=78%.*

## 4.2 Adversarial Robustness Decay

Adversarial accuracy declined monotonically with increasing perturbation magnitude (Figure 2). At ε=0.021 - the smallest non-zero epsilon tested, corresponding to a maximum per-pixel intensity change of 2.1% - accuracy collapsed from 89.3% to 62.0% [95% CI: 54.2-69.8%], representing a 27.3 percentage point drop. By ε=0.150, accuracy had declined to 43.3% [95% CI: 35.4-51.3%], approaching random chance (33.3%). At ε=0.300, accuracy was 34.7% [95% CI: 27.1-42.3%], indistinguishable from random performance. The complete robustness decay profile is presented in Table 1.

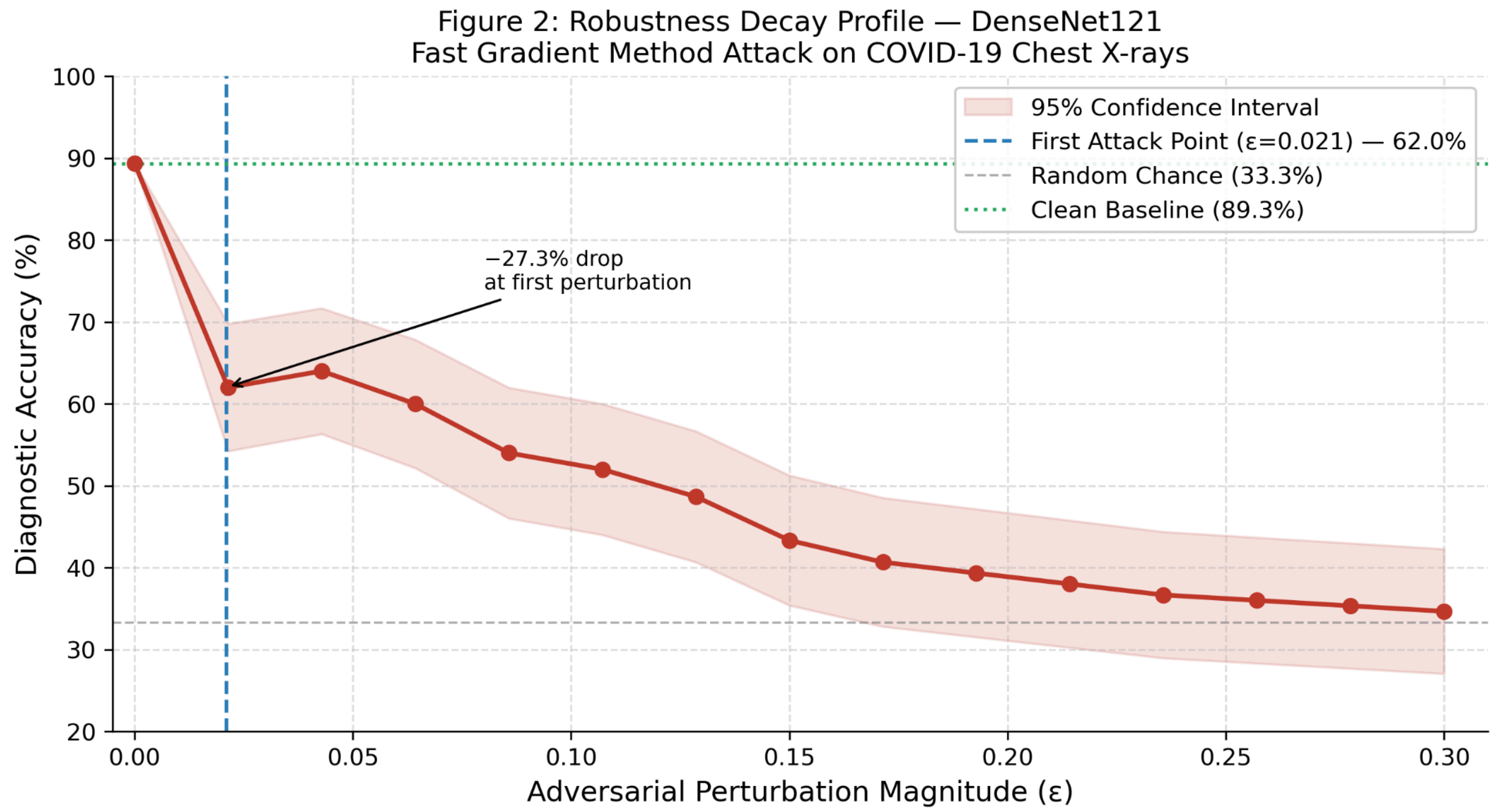

*Figure 2. Robustness decay profile. Diagnostic accuracy (y-axis) plotted against FGM perturbation magnitude ε (x-axis) across 15 levels. Shaded region represents 95% confidence intervals. Blue dashed line: first attack point (ε=0.021). Green dotted line: clean baseline (89.3%). Grey dashed line: random chance (33.3%).*

**Table 1. Robustness Decay - Diagnostic Accuracy Under FGM Attack (DenseNet121, N=150)**

| Epsilon (ε) | Accuracy (%) | 95% CI Lower (%) | 95% CI Upper (%) | Clinical Interpretation |
|---|---|---|---|---|
| 0.000 | 89.3 | 89.3 | 89.3 | Clean baseline |

| Epsilon (ε) | Accuracy (%) | 95% CI Lower (%) | 95% CI Upper (%) | Clinical Interpretation |
|---|---|---|---|---|
| 0.021 | 62.0 | 54.2 | 69.8 | First attack - clinical danger zone |
| 0.043 | 64.0 | 56.3 | 71.7 | Sustained collapse |
| 0.064 | 60.0 | 52.2 | 67.8 | Continued degradation |
| 0.086 | 54.0 | 46.0 | 62.0 | Below 60% threshold |
| 0.107 | 52.0 | 44.0 | 60.0 | Near random chance |
| 0.150 | 43.3 | 35.4 | 51.3 | Approaching random |
| 0.214 | 38.0 | 30.2 | 45.8 | Effectively random |
| 0.300 | 34.7 | 27.1 | 42.3 | Complete collapse |

***Selected epsilon levels shown. Full 15-level results available in supplementary data (exp1_robustness_decay.csv).***

### 4.3 Per-Class Diagnostic Collapse

Per-class analysis at ε=0.021 revealed differential vulnerability across diagnostic classes (Figure 3). Normal patients experienced the greatest accuracy drop (+36.0%), from 78.0% to 42.0% - approaching random chance for a three-class problem. COVID-19 detection collapsed from 100.0% to 72.0% (+28.0%), while Non-COVID Pneumonia accuracy fell from 90.0% to 70.0% (+20.0%). The adversarial confusion matrix revealed that 24 Normal patients were misclassified as Non-COVID Pneumonia and 13 COVID-19 patients were misclassified as Non-COVID Pneumonia - patterns with direct clinical risk implications.

Figure 3: Per-Class Diagnostic Collapse Under Adversarial Attack

*Figure 3. Per-class diagnostic collapse. (a) Grouped bar chart comparing clean vs. adversarial per-class accuracy at ε=0.021. (b) Adversarial confusion matrix showing misclassification patterns. Most dangerous finding: 24 Normal patients predicted as Non-COVID Pneumonia (false positive), and 13 COVID-19 patients misclassified as Non-COVID Pneumonia (missed diagnosis).*

### 4.4 Mitigation Stress Test

All three defensive strategies tested at ε=0.021 failed to restore clinical safety (Figure 4). Gaussian Smoothing (σ=1.0) yielded an accuracy of 34.7% - worse than the undefended adversarial baseline of 61.3% - demonstrating that spatial blurring actively destroys the diagnostic features required for correct classification. Ensemble Voting across five augmentations achieved 72.7%, representing partial recovery but remaining 16.6 percentage points below the clean baseline. Results are summarised in Table 2.

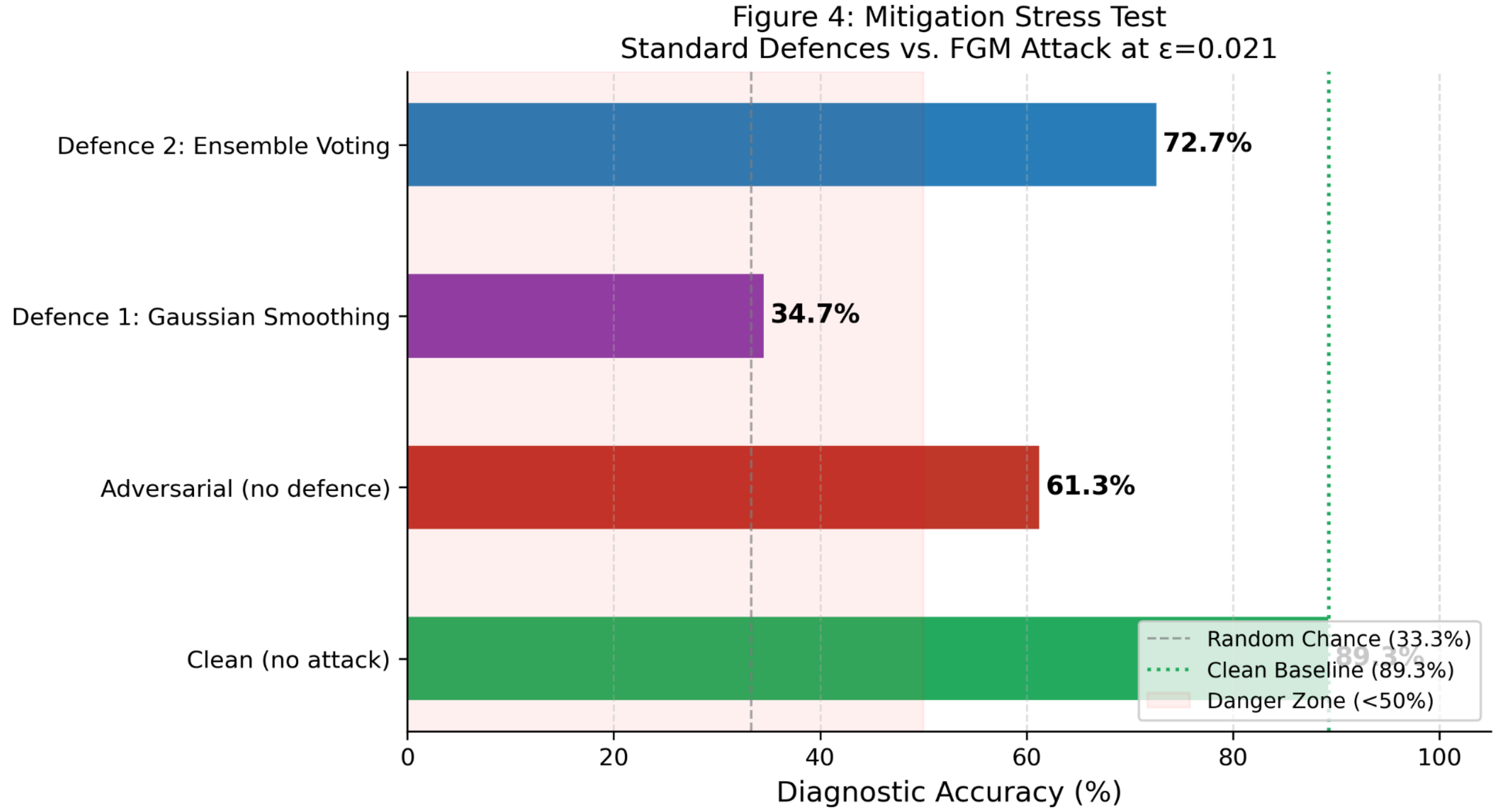

*Figure 4. Mitigation stress test at ε=0.021. Horizontal bar chart comparing clean baseline, undefended adversarial, and three defensive conditions. Red shaded region indicates danger zone (<50% accuracy). Gaussian smoothing performed worse than no defence.*

**Table 2. Mitigation Stress Test Results (ε=0.021, N=150)**

| Condition | Accuracy (%) | vs. Clean Baseline | Clinical Assessment |
|---|---|---|---|
| Clean (no attack) | 89.3 | - | Acceptable |
| Adversarial - no defence | 61.3 | -28.0% | Clinically dangerous |
| Defence 1: Gaussian Smoothing | 34.7 | -54.6% | Worse than no defence |
| Defence 2: Ensemble Voting | 72.7 | -16.6% | Partial recovery - insufficient |

## 4.5 Adversarial Perturbation Visualisation

Visual inspection of clean versus adversarial chest X-rays at ε=0.021 confirmed that perturbations are imperceptible to the human eye (Figure 5). The perturbation map (amplified ×10 for visibility) reveals structured noise concentrated in anatomically informative regions. This structural targeting explains the mechanism of collapse: the adversarial signal selectively disrupts the texture features most discriminative for COVID-19 classification.

Figure 5: Adversarial Perturbations on Clinical Chest X-rays
Perturbation is invisible to the human eye yet causes misdiagnosis

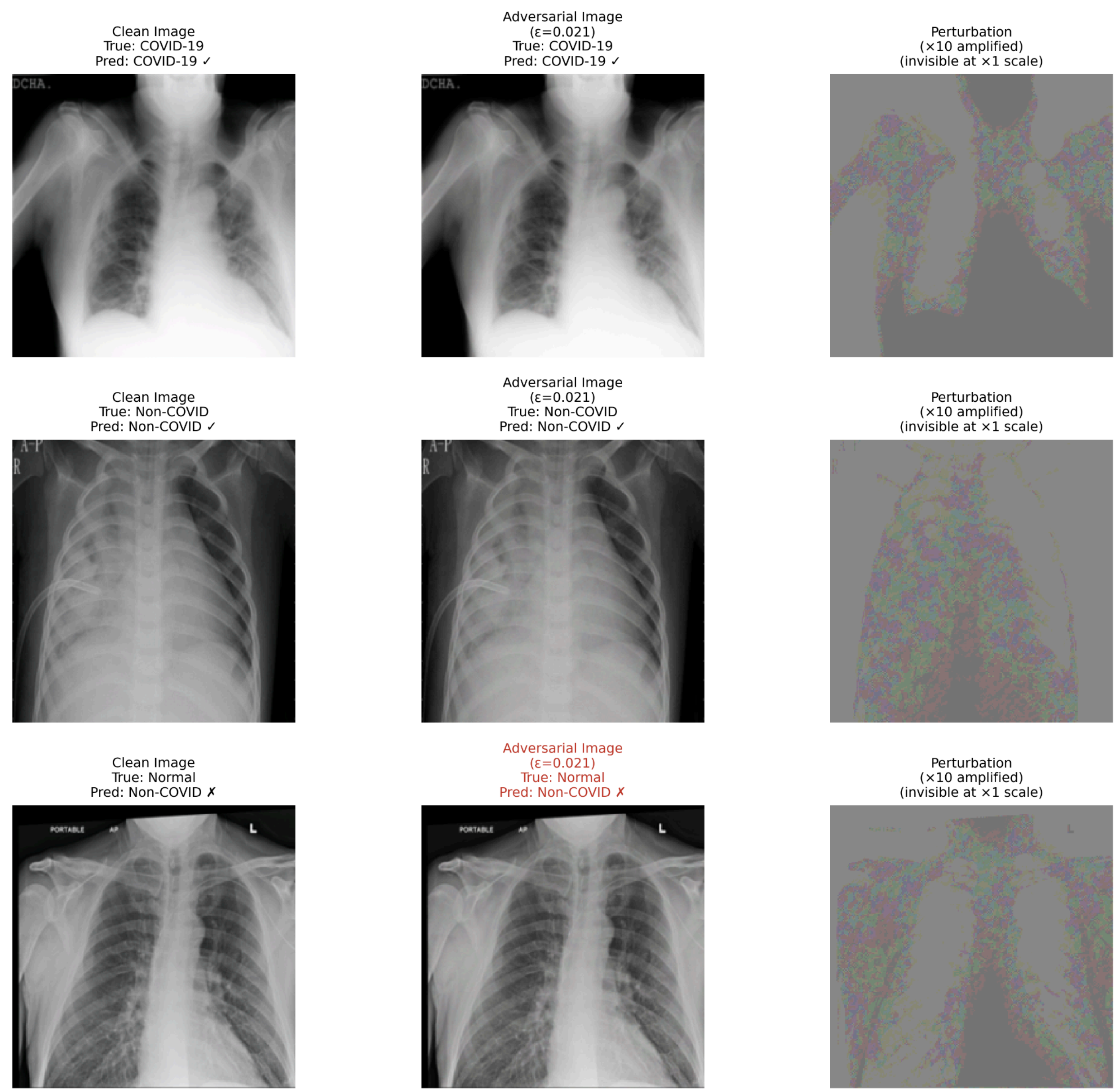

*Figure 5. Adversarial perturbation visualisation across all three diagnostic classes. Left column: clean images. Centre column: adversarial images (ε=0.021) - visually identical to clean. Right column: perturbation map amplified ×10. Red text indicates misclassification. Perturbations are invisible at ×1 scale yet sufficient to cause diagnostic collapse.*

### 4.6 Language Fragility - Llama3.1:8b

Llama3.1:8b demonstrated a systematic accuracy gradient across language registers (Figure 6). Standard English accuracy was 80.0%; Nigerian Pidgin accuracy fell to 65.0% (-15.0 percentage points); Yoruba-inflected English accuracy declined further to 60.0% (-20.0 percentage points). Diagnosis consistency with English predictions was 85.0% for Pidgin and 80.0% for Yoruba-inflected English, indicating that 15-20% of cases received different diagnoses solely due to language register. Five cases exhibited diagnosis flips across languages, with Case 19 (Normal patient) misdiagnosed as Non-COVID Pneumonia in both Pidgin and Yoruba-inflected English.

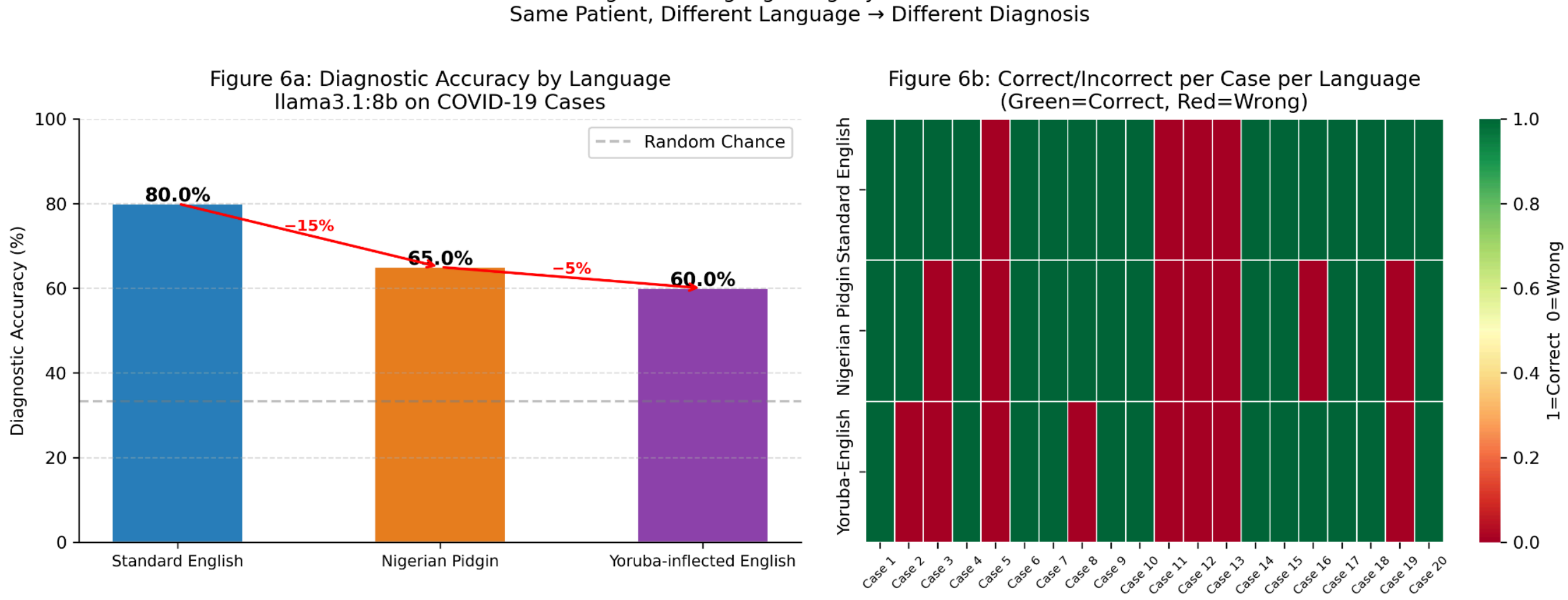

*Figure 6. Llama3.1:8b language fragility. (a) Accuracy by language register with drop annotations. (b) Per-case correct/incorrect heatmap across 20 clinical cases and 3 languages. Green=correct, Red=incorrect.*

### 4.7 Language Fragility - NatLAS

NatLAS exhibited a paradoxical fragility pattern (Figure 7). While achieving higher Standard English accuracy than Llama3.1:8b (85.0% vs 80.0%), it demonstrated substantially greater degradation on Nigerian Pidgin (55.0%, -30.0 percentage points), with diagnosis consistency collapsing to 50.0% - meaning the model produced a different diagnosis for the same patient in 1 of every 2 Pidgin cases compared to its own English prediction. Performance partially recovered on Yoruba-inflected English (75.0%), suggesting differential sensitivity to Pidgin versus Yoruba linguistic features.

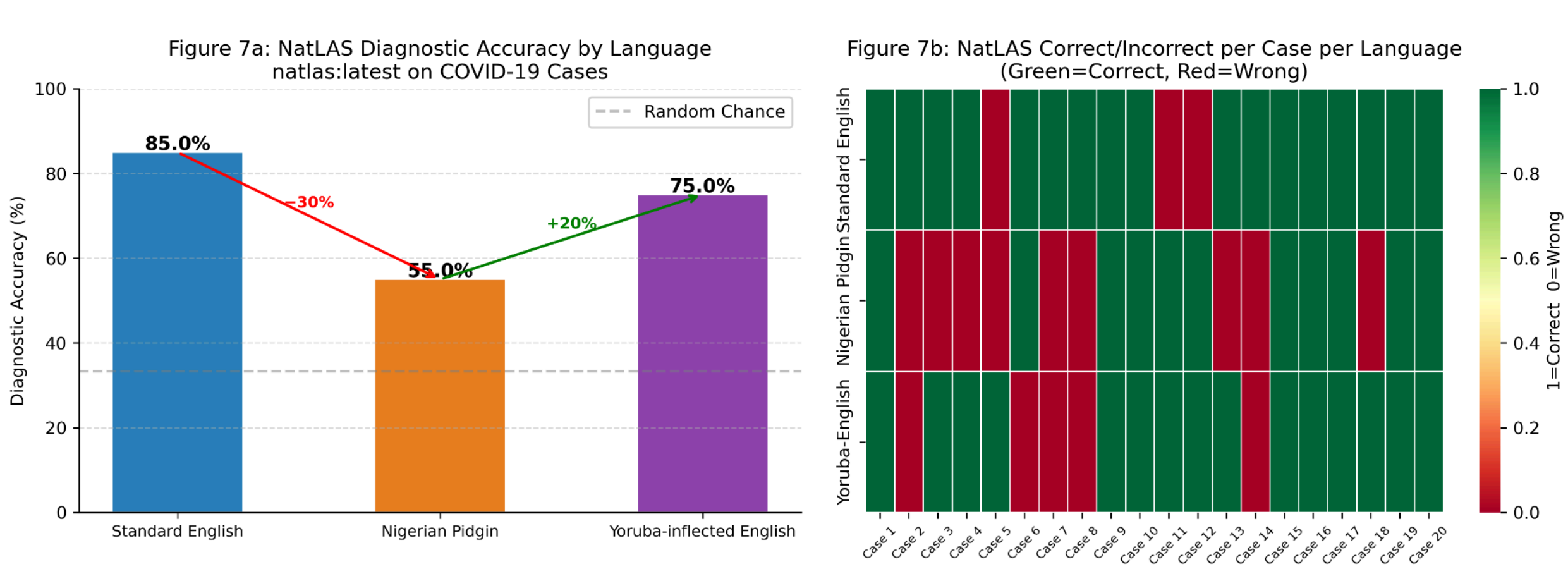

*Figure 7. NatLAS language fragility. (a) Accuracy by language register - note the V-shaped profile with severe Pidgin collapse. (b) Per-case heatmap showing substantially more red cells in the Nigerian Pidgin row compared to Figure 6.*

## 4.8 Head-to-Head Model Comparison

Direct comparison of both models across language registers revealed distinct fragility profiles (Figure 8, Table 3). Llama3.1:8b demonstrated more consistent degradation across languages (monotonic decline), while NatLAS showed severe Pidgin-specific collapse followed by partial Yoruba-English recovery. In terms of consistency with Standard English predictions, Llama3.1:8b maintained higher consistency across both non-English registers (85% Pidgin, 80% Yoruba-English) compared to NatLAS (50% Pidgin, 60% Yoruba-English).

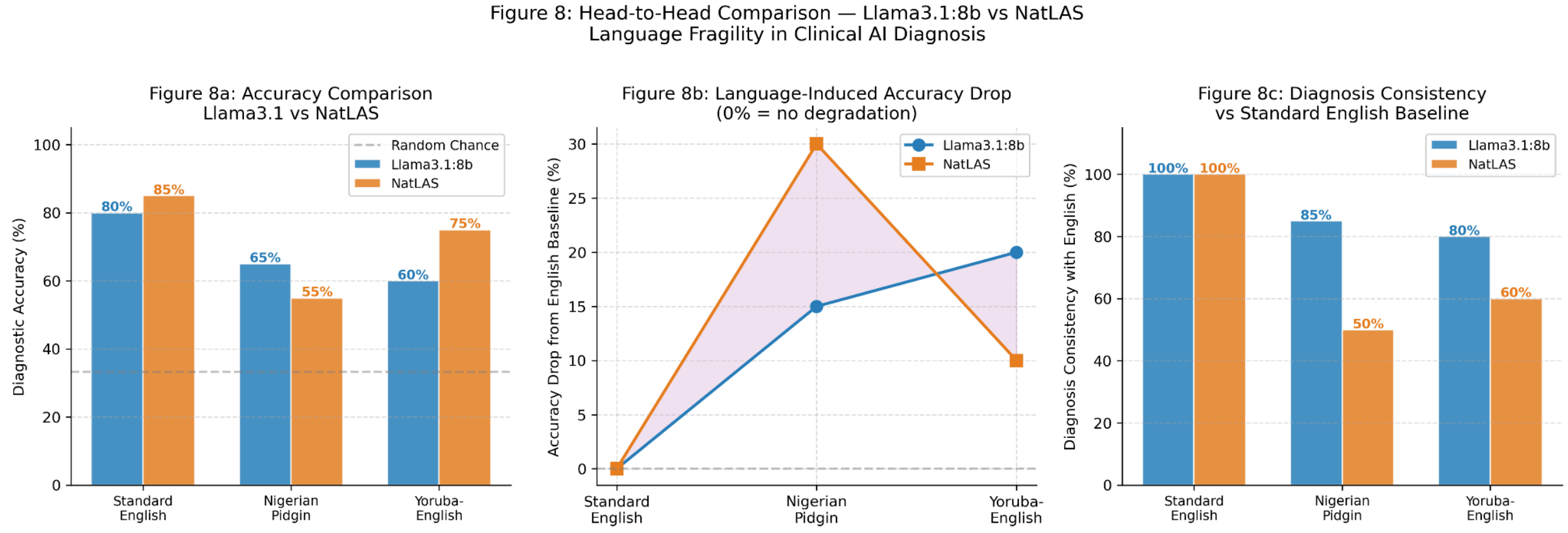

*Figure 8. Head-to-head model comparison. (a) Accuracy comparison grouped by language register. (b) Language-induced accuracy drop from English baseline - NatLAS shows steeper Pidgin collapse. (c) Diagnosis consistency with Standard English - NatLAS consistency drops to 50% on Pidgin.*

**Table 3. Language Fragility Summary - Llama3.1:8b vs NatLAS (N=20 cases per language)**

| Language Register | Llama3.1:8b Accuracy | NatLAS Accuracy | Llama Consistency | NatLAS Consistency |
| --- | --- | --- | --- | --- |
| Standard English | 80.0% | 85.0% | 100% | 100% |
| Nigerian Pidgin | 65.0% | 55.0% | 85% | 50% |
| Yoruba-inflected English | 60.0% | 75.0% | 80% | 60% |

## 5. Discussion

### 5.1 The Adversarial Failure Envelope

Our findings establish that DenseNet121 - the architecture underlying widely deployed clinical AI systems - possesses a critical failure envelope at $\varepsilon=0.021$ under FGM attack. This epsilon value, corresponding to a maximum per-pixel intensity change of 2.1% of the full intensity range, falls well within the range of natural digital artefacts encountered in clinical settings: scanner calibration differences, JPEG compression artefacts, and network transmission degradation all introduce comparable perturbation magnitudes [6,7]. The implication is that current clinical AI systems may already be operating in their adversarial danger zone under routine deployment conditions.

The finding that Gaussian smoothing (34.7%) performed worse than no defence (61.3%) is particularly significant. This result indicates that spatial blurring destroys the fine-grained textural features upon which DenseNet121 relies for COVID-19 classification, without eliminating the adversarial signal - leaving the model with degraded inputs and no compensatory mechanism [3]. This contradicts naive intuitions about image preprocessing as a defensive strategy and aligns with theoretical analyses showing that smoothing-based defences are provably ineffective against gradient-based attacks [23].

The observation that Normal patients experienced the greatest per-class accuracy drop (42.0% at $\varepsilon$=0.021) is clinically the most alarming finding. False positive diagnoses - misclassifying healthy patients as having COVID-19 or pneumonia - generate unnecessary treatment burden, psychological distress, and resource misallocation. In PHC settings already operating under resource constraints, this failure mode could overwhelm triage workflows.

### 5.2 The Localisation Paradox in NatLAS

The counter-intuitive finding that NatLAS - explicitly designed for African linguistic contexts - exhibited greater Pidgin fragility than the general-purpose Llama3.1:8b constitutes what we term the 'Localisation Paradox.' We hypothesise that this occurs because NatLAS was trained primarily on formal written registers of Nigerian languages (including WAPE - the BBC variety of Pidgin) rather than the conversational community-level Naija spoken in PHC interactions [14,17]. This distinction, documented in the WARRI benchmark [14], shows a 24.3 ChrF++ point gap between WAPE and community Naija performance even in models explicitly targeting African languages.

The 50% diagnosis consistency for NatLAS on Pidgin - meaning the model produces different diagnoses for identical patients depending solely on language register - represents a form of 'algorithmic inequity' where patients who communicate in their native dialect receive systematically different clinical assessments [22]. This is a form of digital redlining that requires deliberate intervention through diverse training datasets and culturally relevant stakeholder panels [35].

### 5.3 Governance and Regulatory Implications

The vulnerabilities identified in this audit occur in a regulatory vacuum. No framework currently mandates adversarial robustness testing or cross-lingual fairness evaluation before clinical AI deployment in Nigeria or most of sub-Saharan Africa [2,33]. The EU AI Act and NIST AI RMF both classify medical AI as high-stakes, requiring rigorous documentation and monitoring - standards that have not yet been translated into the Nigerian regulatory context [34]. Our findings provide empirical grounding for the following minimum requirements: (1) all clinical AI must publish an Adversarial Robustness Map defining its failure envelope; (2) models must be evaluated against community-level Naija corpora in addition to Standard English; and (3) deployment dashboards must include disaggregated accuracy metrics by language register and demographic subgroup [35].

### 5.4 Limitations

Several limitations constrain the generalisability of our findings. First, our adversarial experiments were conducted in a white-box threat model, assuming attacker access to model gradients - a condition that may overestimate real-world risk. Second, the language fragility experiment used 20 authored clinical vignettes rather than real patient records, which may not fully capture the syntactic and lexical diversity of authentic PHC presentations. Third, fine-tuning was conducted on 200 images per class, which may introduce overfitting not present in large-scale trained clinical models. Fourth, all experiments were conducted on a single consumer device (Apple Silicon MacBook, 16GB RAM), limiting batch sizes and the number of test images relative to the full dataset. Future work should validate these findings using real patient records, larger test cohorts, and black-box threat models.

## 6. Conclusions

This study presents the first systematic dual audit of adversarial image fragility and cross-lingual diagnostic drift in clinical AI, contextualised for Primary Health Centre deployment in Nigeria. Our key conclusions are:

1. Brittle Foundations: DenseNet121 collapses from 89.3% to 62.0% diagnostic accuracy at $\varepsilon=0.021$ - a perturbation invisible to the human eye and within the range of natural digital artefacts.
2. Defence Failure: Standard defensive strategies - Gaussian smoothing and ensemble voting - fail to restore clinical safety, indicating that adversarial vulnerability is architecturally embedded.

3. Language Equity Gap: Both Llama3.1:8b and NatLAS degrade significantly on Nigerian Pidgin and Yoruba-inflected English, with NatLAS exhibiting a paradoxically greater Pidgin collapse (30%) despite African-context training.
4. Mandatory Auditing: No clinical AI system should be deployed without a published Adversarial Robustness Map and Cross-Lingual Drift Profile.
5. Sovereign Solutions: Safe clinical AI for African PHC settings requires adversarially hardened architectures trained on community-level Naija corpora - not just formal Nigerian English.

## Data Availability

All experimental code, figures, and data tables are publicly available at: https://github.com/anthoniooladimeji11-coder/clinical-ai-safety-audit. The COVID-QU-Ex dataset is available under CC-BY-SA-4.0 at https://www.kaggle.com/datasets/anasmohammedtahir/covidqu. Experiments are fully reproducible using the provided Jupyter notebook on standard consumer hardware.

## Competing Interests

The authors declare no competing interests.

## References

1. Wahl B. et al. Artificial intelligence (AI) and global health: how can AI contribute to health in resource-poor settings? BMJ Global Health, 3(4), e000798 (2018). https://doi.org/10.1136/bmjgh-2018-000798

2. Okafor C. et al. The utilization of artificial intelligence (AI) and machine learning (ML) for health in Nigeria: a rapid review. Journal of Medical Artificial Intelligence (2024). https://jmai.amegroups.org/article/view/11267

3. Amgad M. et al. Robust and Interpretable Chest X-ray Classification via Diffusion Purification and Concept-Based Adversarial Detection. Journal of Object Technology in Biomedical Research, 2025. https://doi.org/10.1016/j.media.2025.103375

4. Tahir A.M. et al. COVID-19 infection localization and severity grading from chest X-ray images. Computers in Biology and Medicine, 139, 105002 (2021). https://doi.org/10.1016/j.compbiomed.2021.105002

5. Adeyemi O. et al. WeCAViT: A Weighted CNN-ViT model for Pneumonia Detection in Chest X-rays. IEEE Access, 2025. https://www.researchgate.net/publication/389527548

6. Rahman T. et al. An enhanced ensemble defense framework for boosting adversarial robustness of intrusion detection systems. Expert Systems with Applications, 2025. https://doi.org/10.1016/j.eswa.2025.126800

7. Kaviani S. et al. Adversarial Robustness of Deep Learning in Medical Imaging: A Comprehensive Survey and Benchmark. International Journal of Advanced Computer Science and Applications (IJACSA), 16(12) (2025). https://thesai.org/Publications/ViewPaper?Volume=16&Issue=12&Code=ijacsa&SerialNo=78

8. Srinivasan A., Sritharan D.V., Chadha S., Fu D., Hossain O., Breuer G.A., and Aneja S. Adversarial Robustness of Capsule Networks for Medical Image Classification. medRxiv (2026). https://doi.org/10.64898/2026.03.09.26347900

9. Ucar F. and Korkmaz D. COVIDiagnosis-Net: Deep Bayes-SqueezeNet based diagnosis of the coronavirus disease 2019 (COVID-19) from X-ray images. Medical Hypotheses, 140, 109761 (2020). https://doi.org/10.1016/j.mehy.2020.109761

10. Rajpurkar P. et al. CheXNet: Radiologist-Level Pneumonia Detection on Chest X-Rays with Deep Learning. arXiv:1711.05225 (2017). https://arxiv.org/abs/1711.05225

11. Deng J. et al. ImageNet: A large-scale hierarchical image database. In Proceedings of IEEE CVPR, 248-255 (2009). https://doi.org/10.1109/CVPR.2009.5206848

12. Goodfellow I.J., Shlens J., and Szegedy C. Explaining and Harnessing Adversarial Examples. In International Conference on Learning Representations (ICLR) (2015). arXiv:1412.6572

13. Nicolae M.I. et al. Adversarial Robustness Toolbox v1.0.0. arXiv:1807.01069 (2019). https://arxiv.org/abs/1807.01069

14. Adelani D.I., Dogruoz A.S., and Aremu A.K. Does Generative AI speak Nigerian Pidgin? Issues about Representativeness and Bias for Multilingualism in LLMs. In Findings of NAACL 2025. ACL Anthology (2025). arXiv:2404.19442

15. Nekoto W. et al. Participatory Research for Low-resourced Machine Translation: A Case Study in African Languages. In Findings of EMNLP 2020. ACL Anthology. https://aclanthology.org/2020.findings-emnlp.195

16. Bender E.M., Gebru T., McMillan-Major A., and Shmitchell S. On the Dangers of Stochastic Parrots: Can Language Models Be Too Big? In Proceedings of FAccT 2021, 610-623. https://doi.org/10.1145/3442188.3445922

17. Coggins W., McKenzie J., Youm S., Mummaleti P., Gilbert J., Ragan E., and Dorr B.J. That Ain't Right: Assessing LLM Performance on QA in African American and West African English Dialects. In Proceedings of the 9th Widening NLP Workshop (WiNLP), ACL 2025. https://aclanthology.org/2025.winlp-main/

18. Ogunremi T. et al. N-ATLAS: Nigerian Atlas for Languages and AI at Scale. arXiv:2509.08234 (2025). https://arxiv.org/abs/2509.08234

19. Garnerin M. et al. Google Fleurs: Few-shot Learning Evaluation of Universal Representations of Speech. In IEEE Spoken Language Technology Workshop (2022). https://doi.org/10.1109/SLT54892.2023.10022793

20. Masakhane. Participatory Research for Low-resourced Machine Translation: Community Approaches to African Language AI. Masakhane White Paper (2020). https://www.masakhane.io

21. World Health Organization. Integrated Management of Childhood Illness (IMCI): Chart Booklet. WHO Press, Geneva (2014). https://www.who.int/publications/i/item/9789241506823

22. Fleisig G. et al. When the Majority is the Minority: Cross-lingual Learning in Low-resource Settings. In Proceedings of ACL 2023. https://aclanthology.org/2023.acl-long.77

23. Cheng Z., Yang J., Dai W., and Sun J. Provable Defense Framework for LLM Jailbreaks via Noise-Augumented Alignment. arXiv:2602.01587 (2026). https://arxiv.org/abs/2602.01587

24. Huang G., Liu Z., van der Maaten L., and Weinberger K.Q. Densely Connected Convolutional Networks. In Proceedings of IEEE CVPR, 4700-4708 (2017). https://doi.org/10.1109/CVPR.2017.243

25. Li H. et al. Adaptive noise-augmented attention for enhancing Transformer fine-tuning on longitudinal medical data. Frontiers in Artificial Intelligence, 8, 1663484 (2025). https://doi.org/10.3389/frai.2025.1663484

26. World Health Organization. WHO Guidelines for Malaria. WHO Press, Geneva (2025). https://www.who.int/publications/i/item/guidelines-for-malaria

27. Evans L. et al. Surviving Sepsis Campaign: International Guidelines for Management of Sepsis and Septic Shock 2021. Intensive Care Medicine, 47, 1181-1247 (2021). https://doi.org/10.1007/s00134-021-06506-y

28. Federal Ministry of Health Nigeria. Standard Treatment Guidelines (5th Edition). Federal Ministry of Health, Abuja (2022). https://www.health.gov.ng

29. Wu Y. et al. Uncertainty-aware feature-weighted ensemble framework for heart disease prediction. PMC — PLOS ONE (2025). https://pmc.ncbi.nlm.nih.gov/articles/PMC13106842/

30. Zhang X. et al. LISArD: learning image similarity to defend against gray-box adversarial attacks. PeerJ Computer Science, e3735 (2025). https://doi.org/10.7717/peerj-cs.3735

31. Xia S., Ding M., Kong C., Yang W., and Jiang X. Feature-Space Adversarial Robustness Certification for Multimodal Large Language Models. arXiv:2601.16200 (2026). https://arxiv.org/abs/2601.16200

32. Liu Z. et al. A ConvNet for the 2020s. In Proceedings of IEEE CVPR, 11976-11986 (2022). https://doi.org/10.1109/CVPR52688.2022.01167

33. Paradigm Initiative. LONDA 2025 Digital Rights and Inclusion in Africa Report. Paradigm Initiative Press (2026). https://paradigmhq.org/wp-content/uploads/2026/04/LONDA-2025-REPORT-2.pdf

34. Challen R. et al. Artificial intelligence, bias and clinical safety. BMJ Quality and Safety, 28(3), 231-237 (2019). https://doi.org/10.1136/bmjqs-2018-008370

35. Topol E.J. Artificial Intelligence in Healthcare: A Narrative Review of Recent Clinical Applications, Implementation Strategies, and Challenges. PMC — npj Digital Medicine (2025). https://pmc.ncbi.nlm.nih.gov/articles/PMC12764347/